\documentclass[a4paper,10pt]{article}
\usepackage{latexsym, amsmath, amsfonts, amssymb, array, dcolumn, hhline, longtable, enumerate,psfrag,bm}
\usepackage[english]{babel}
\usepackage[dvips]{graphicx}
\usepackage{epsfig}
\usepackage[small,bf]{caption}
\begin{document}
%
%
%

\noindent {\bf \Large{Modeling of graphite oxide}}

\vspace{0.5 cm}

\noindent
{\bf \large {D.W. Boukhvalov and M. I. Katsnelson}}

\vspace{0.5 cm}

\noindent {\it Institute for Molecules and Materials, Radboud
University Nijmegen, Heyendaalseweg 135, 6525 AJ Nijmegen,
~~~~~~~~~ ~~~~~~~~~ ~~~~~The Netherlands}

\begin{abstract}
Based on density functional calculations, optimized structures of
graphite oxide are found for various coverage by oxygen and
hydroxyl groups, as well as their ratio corresponding to the
minimum of total energy. The model proposed describes well known
experimental results. In particular, it explains why it is so
difficult to reduce the graphite oxide up to pure graphene.
Evolution of the electronic structure of graphite oxide with the
coverage change is investigated.
\end{abstract}



\section{Introduction}

Despite the fact that graphite oxide (GO) was first derived more
than a century ago$^{1,2}$ its structure and
chemical composition remains not quite clear yet. GO can be used
for production of graphite nanoparticles and an insulating
material for nanodevices$^{3,4}$. Recently, after
discovery of extraordinary electronic properties of single-layer
carbon, graphene$^{5-8}$, and successful exfoliation of
layers in GO$^{9,10}$ it is considered as a perspective source
of a ``cheap graphene''$^{3}$. Direct structural
information about GO can be hardly obtained (for the most of
structural methods the use of bulk crystals are desirable whereas
the GO exists mainly in solutions) which makes especially important
theoretical modeling of its structure and properties.

Original methods of preparation of GO have been modified
afterwards$^{11-14}$ which allows to vary a bit
its chemical composition. In Nakajima and Matsuo paper$^{11}$
the chemical compositions of GO derived by the methods developed
by Brodie$^{1}$ and Staudenmaier$^{2}$
were determined as C$_8$H$_{2.54}$O$_{3.91}$ and
C$_8$H$_{4.61}$O$_{6.70}$, respectively. This means that both
hydroxyl groups connected with single carbon atoms and oxygen
atoms connected with two carbon atoms present in GO (see Fig.
\ref{fig1}). According to Szab\'o and co-workers$^{12}$,
the chemical composition of different samples of GO varies from
C$_8$H$_{1.20}$O$_{3.12}$ to C$_8$H$_{1.60}$O$_{3.92}$, according
to Hontora-Lucas et al. paper$^{13}$ - from C$_8$(OH)$_{1.38}$O$_{0.63}$
to C$_8$(OH)$_{1.64}$O$_{0.79}$, and, according to Cassagneau
and co-workers$^{15}$ -
from C$_{12}$HO$_{2}$ to C$_{15}$H$_{3}$O$_{4}$. In general, one
can conclude that the chemical composition of GO, not considering
groups coupled with graphene edges, varies in a range from
C$_8$H$_{2}$O$_{3}$ to C$_8$H$_{4}$O$_{5}$. It is a common opinion
(see the papers cited above) that oxygen in GO mainly presents in
hydroxyl groups or in bridges connecting two carbon atoms in
graphene layers whereas the amount of carboxyl, as well as
carbonyl, groups is relatively small. Thus, the two limiting
compositions of GO can be presented as C$_8$(OH)$_{2}$ and
C$_8$(OH)$_{4}$O. All chemical formulas of GO obtained
experimentally manifest the coverage of graphene between 25\% and
75\%, which means that, at least, quarter of carbon-carbon bonds
in graphene layer are double bonds whereas the rest is single
bonds like in diamond. Indeed, both XPS$^{13,16}$ and
infrared$^{17}$ spectroscopy data confirm coexistence of $sp^3$
and $sp^2$ electron configurations of carbon.

Based on experimental data$^{11}$ and additional
measurement results, a model of GO has been suggested
by Nakajima and co-workers$^{18}$.
According to this model, the GO structure is
intermediate between two ideal structures, C$_8$O$_2$ and
C$_8$(OH)$_4$ (Fig. \ref{fig2}c and h respectively). Later models$^{19,20}$
differ mainly by assumptions concerning edge
groups. As a result, the GO structure is described as a
combination of completely covered and completely uncovered stripes
of graphene which is confirmed also by recent theoretical results$^{21,22}$.
Recently, mechanical properties of GO have been
simulated, based on modeling of nanoobjects functionalized by
oxygen from one side$^{23}$ or on experimental data on
chemical composition of GO$^{24}$.

However, due to the stripes of uncovered graphene, GO should be
conducting, according to these models. At the same time,
experimentally GO becomes conducting only after a very strong
reduction$^{4,25}$ whereas typically GO is
insulating. Here, based on density functional calculations, we
formulate a model of insulating GO. We investigate also a
transition to conducting state at the reduction
and explain why it is so difficult to clean GO
completely and to derive pure graphene from GO.

\section{Computational Method}

Some general factors determining chemical functionalization of
graphene have been investigated in our previous work$^{26}$
using hydrogenization as an example. First, graphene is a very
flexible material, and a chemisorption of even single hydrogen
atom leads to essential distortions of the graphene sheet with a
radius approximately 5 \AA, and these lattice distortions are of
crucial importance for energetics of the process. Second, for the
chemisorption of two hydrogen atoms the configuration where they
are bonded with two neighboring carbon atoms from opposite sides
of the sheet turns out to be the most energetically favorable.
Third, complete coverage by hydrogen provides the global minimum
of energy. These features are relevant, as we will show here, also
for the case of GO.

We used the pseudopotential density functional SIESTA package for
electronic structure calculations$^{27,28}$ with the
generalized gradient approximation for the density functional$^{29}$,
with energy mesh cutoff 400 Ry, and $k$-point
11$\times$11$\times$1 mesh in Monkhorst-Park scheme$^{30}$.
During the optimization, the electronic ground states was found
self-consistently by using norm-conserving pseudopotentials to
cores and a double-$\zeta$ plus polarization basis of localized
orbitals for carbon and oxygen, and double-$\zeta$ one for
hydrogen. Optimization of the bond lengths and total energies was
performed with an accuracy 0.04 eV /\AA ~and 1 meV, respectively.
This method is frequently used for computations of electronic
structure of graphene$^{26,31-33}$.

To compute the properties of layered GO we have carried out
calculations of the corresponding structures for the case of
25\% coverage by hydroxyl groups. Instead of GGA, we use here the
LDA approximation which is known to be more accurate to describe
interlayer coupling in graphite and other van der Waals systems$^{34}$.
The basis for carbon atoms was optimized as proposed
earlier for pure graphite$^{35}$.

Chemisorption energies were calculated by standard formulas used,
e.g., earlier for the case of chemisorption of hydrogen on
graphene$^{26}$ and solution of carbon in $\gamma$-iron$^{36}$.
Thus, energy of chemisorption of single oxygen atom at eight
carbon atoms (Fig. 2a) is calculated as
E$_{form}$=E$_{C_{8}O}$-E$_{C_8}$-E$_{O_2}$/2 where E$_{C_{8}O}$
is the total energy of the supercell found by self-consistent
calculations after optimization of geometric structure, E$_{C_8}$
is the total energy of graphene supercell, and E$_{O_2}$ is the
energy of oxygen molecule. For the case of hydroxyl group, instead
of oxygen, its energy was calculated with respect to water in
gaseous phase: E$_{OH}$=E$_{H_2 O}$-E$_{H_2}$/2. Alternatively,
the chemisorption energy can be calculated as E$_{OH}$=E$_{H_2
O}$/2+E$_{O_2}$/4. These two expressions estimate the
chemisorption energy from above and from below (see Fig.
\ref{fig3}b where the results corresponding to the first and to
the second expression are shown by dashed green and dotted blue,
respectively). To be specific, in further discussions we will use
the first estimation (the dashed green line). Actually, the
chemisorption energy for GO containing both oxygen and hydroxyl
groups depend on its chemical composition. For example, the
chemisorption energy of oxygen and OH group in the system
C$_8$(OH)$_4$O are E$_{chem}$ = E$_{C_{8}(OH)_4}$ + E$_{O_2}$/2
-  E$_{C_{8}(OH)_{4}O}$ and E$_{chem}$ = (E$_{C_{8}(OH)_{2}O}$ +
2E$_{OH}$ - E$_{C_{8}(OH)_{4}O}$)/2, respectively.

To check an accuracy of the method used we have calculated
formation energy of the water from molecular oxygen and hydrogen
in gaseous phase. We have found the value 213.4 kJ/mol which is
rather close to the experimental value 241.8 kJ/mol.
Underestimation of the energy by approximately 10\% is typical for
GGA calculations$^{36}$. Also, we have calculated equilibrium
interatomic distances for graphene, molecular oxygen, hydrogen,
and water, as well as interlayer distances in graphene. When
drawing the pictures of density of states, a smearing by 0.3 eV
was used.

\section{Results and Discussion}

\begin{figure}[ht]
 \begin{center}
   \centering
\includegraphics[width=5.2 in]{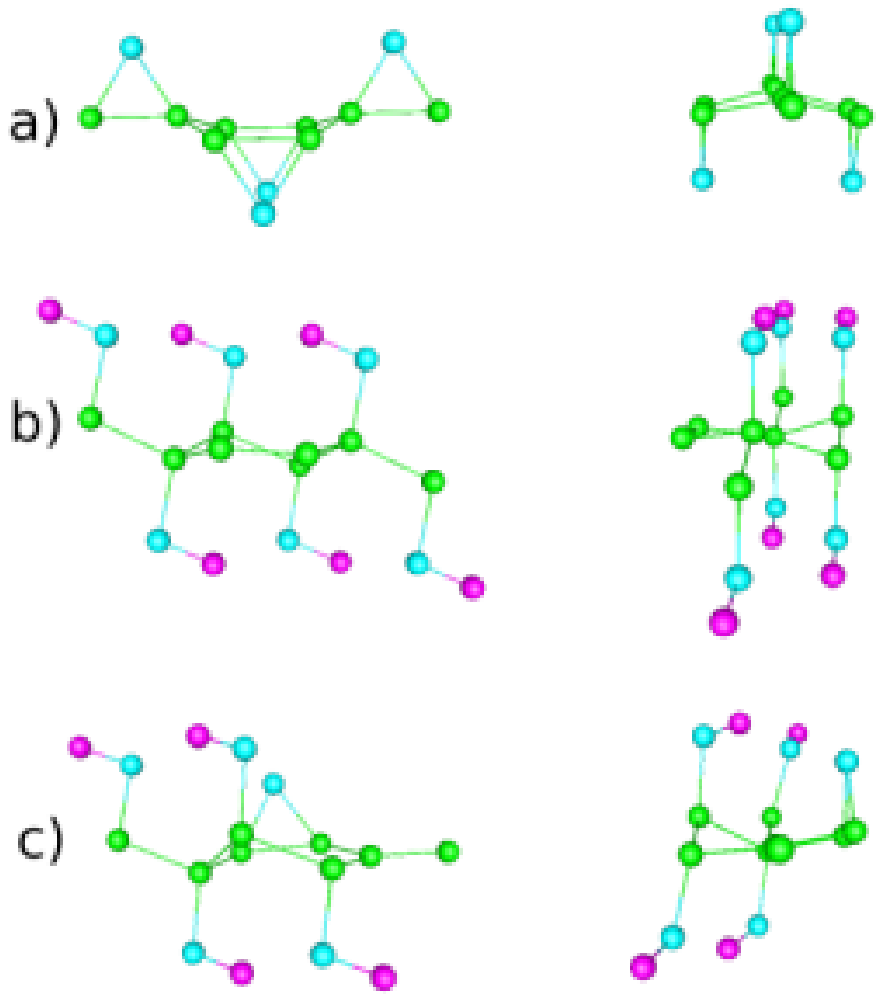}
\caption{The most stable configurations of graphene functionalized
by oxygen only (a), hydroxyl groups only (b), and both oxygen and
hydroxyl groups (c). Carbon, oxygen and hydrogen atoms are shown
in green, blue, and violet, respectively.}
            \label{fig1}
 \end{center}
\end{figure}

\begin{figure}[ht]
 \begin{center}
   \centering
\includegraphics[width=5.2 in]{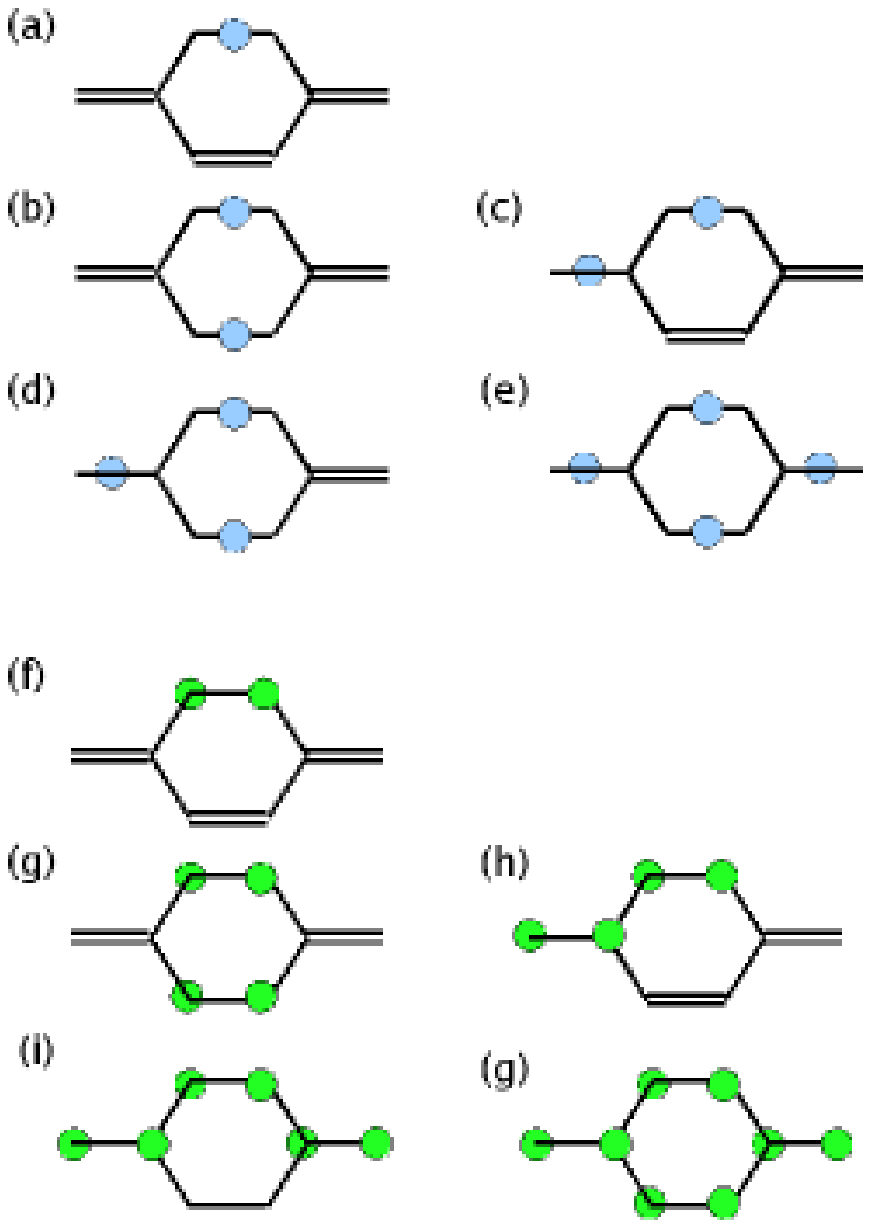}
\caption{A sketch of functionalization of graphene by (a-e)
oxygens (blue circles), and (f-g) hydroxy groups (green circles).}
            \label{fig2}
 \end{center}
\end{figure}

\begin{figure}[ht]
 \begin{center}
   \centering
\includegraphics[width=5.2 in]{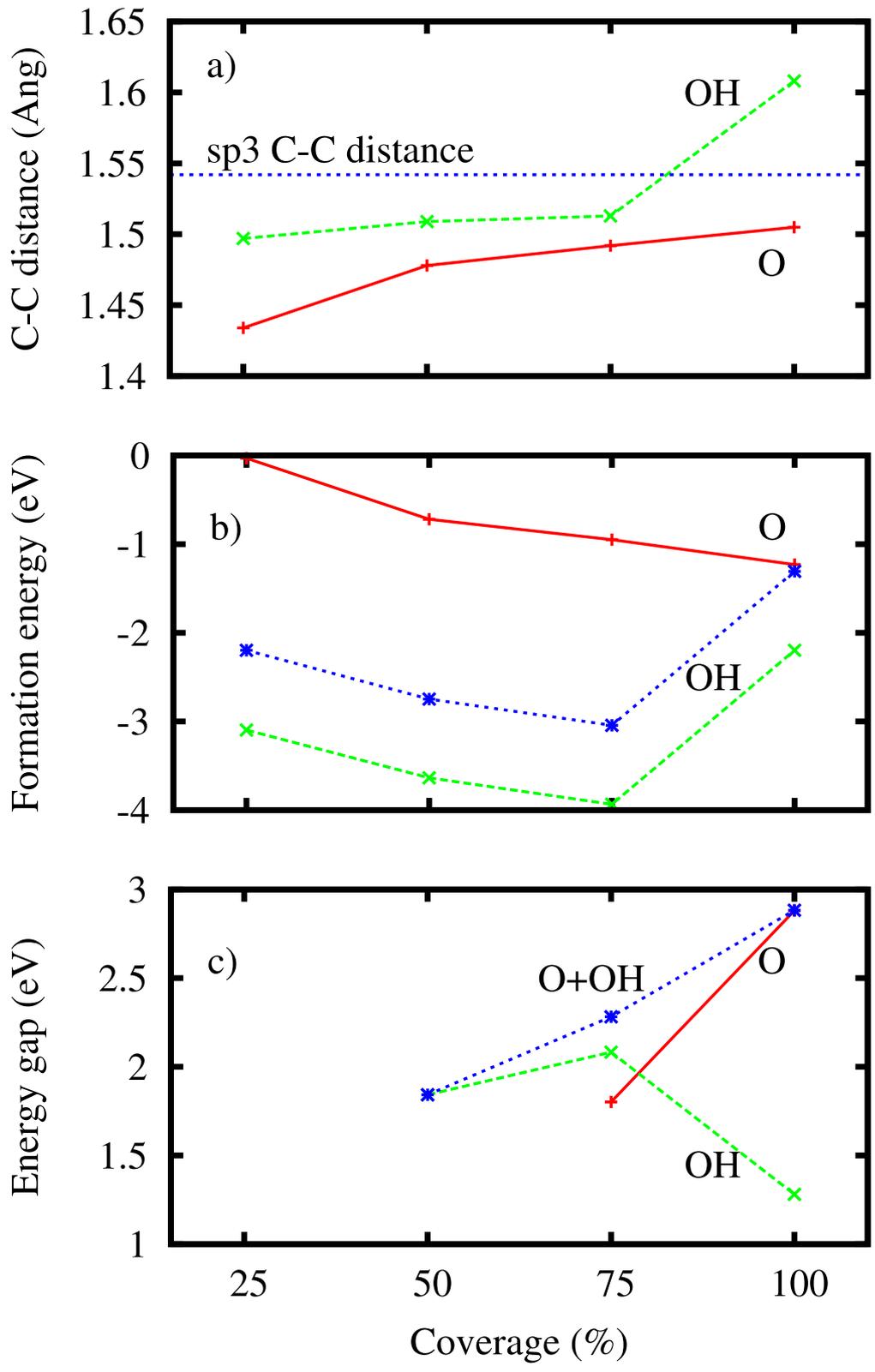}
\caption{Dependence of carbon-carbon bond length (a),
chemisorption energy (b), and electron energy gap (c) on coverage
(see the text).}
            \label{fig3}
 \end{center}
\end{figure}

We start our simulations with the case of oxygen chemisorption,
then consider the chemisorption of hydroxyl groups and, at last,
investigate their various combinations. Let us consider first a
supercell of graphene containing 8 carbon atoms, the chemisorption
of two of them corresponding 25\% coverage. In contrast with
hydrogen, oxygen forms a bridge between two carbon atoms, as shown
in Fig. \ref{fig1}a. As well as for the case of hydrogen, the
chemisorption leads to distortions of graphene sheets when the
atoms coupled to oxygen are shifted up and their neighbors are
moved in the opposite direction. This makes chemisorption of the
next oxygen atom from the opposite side of graphene sheet the most
energetically profitable (Fig. \ref{fig1}a). Various oxygen
configurations for various coverage are sketched in Fig.
\ref{fig2}a-e and the computational results for carbon-carbon
distances, chemisorption energies, and electron energy gaps are
presented in Fig. \ref{fig3}.
One can see from Fig. \ref{fig3}a that the length of
bonds between functionalized carbon atoms grows from the standard
value for graphene, 1.42 \AA, to the standard value for diamond,
1.54 \AA, at the coverage increase which corresponds to the
transition from $sp^2$ to $sp^3$ hybridization of carbon atoms.
The chemisorption energy increases in absolute value with the
coverage increase and the most stable is the configuration
displayed in Fig. \ref{fig1}a. The gap in electron energy spectrum
opens starting form 75\% coverage where its value is 1.8 eV; with
the coverage increase, it grows up to 2.9 eV.

Hydroxyl groups are bonded with graphene similar to hydrogen$^{26}$,
that is, they sit at neighboring carbon atoms from
opposite sides of the graphene sheet (Fig. \ref{fig1}b).
Distortions of the sheet is stronger in this case than in the case
of hydrogen, partially, due to interaction between the hydroxyl
groups leading to ordering of distortion (see Fig. \ref{fig1}b,c).
Various calculated configurations are sketched in Fig.
\ref{fig2}f-g. The chemisorption energy was calculated with
respect to water which is probably the most informative to
consider reduction of GO. With the coverage increase, the
carbon-carbon distance growth, as shown in Fig. \ref{fig3}a. For
the case of complete coverage it turns out to be larger than a
standard one for a single bond ($sp^3$ hybridization) which means
a situation close to break of the graphene sheet. In contrast with
the cases of hydrogen and oxygen, the chemisorption energy is not
monotonous as a function of coverage reaching the minimum at 75\%
which should correspond, therefore, to the most optimal
configuration (Fig. \ref{fig1}b).

\begin{figure}[ht]
 \begin{center}
   \centering
\includegraphics[width=5.2 in]{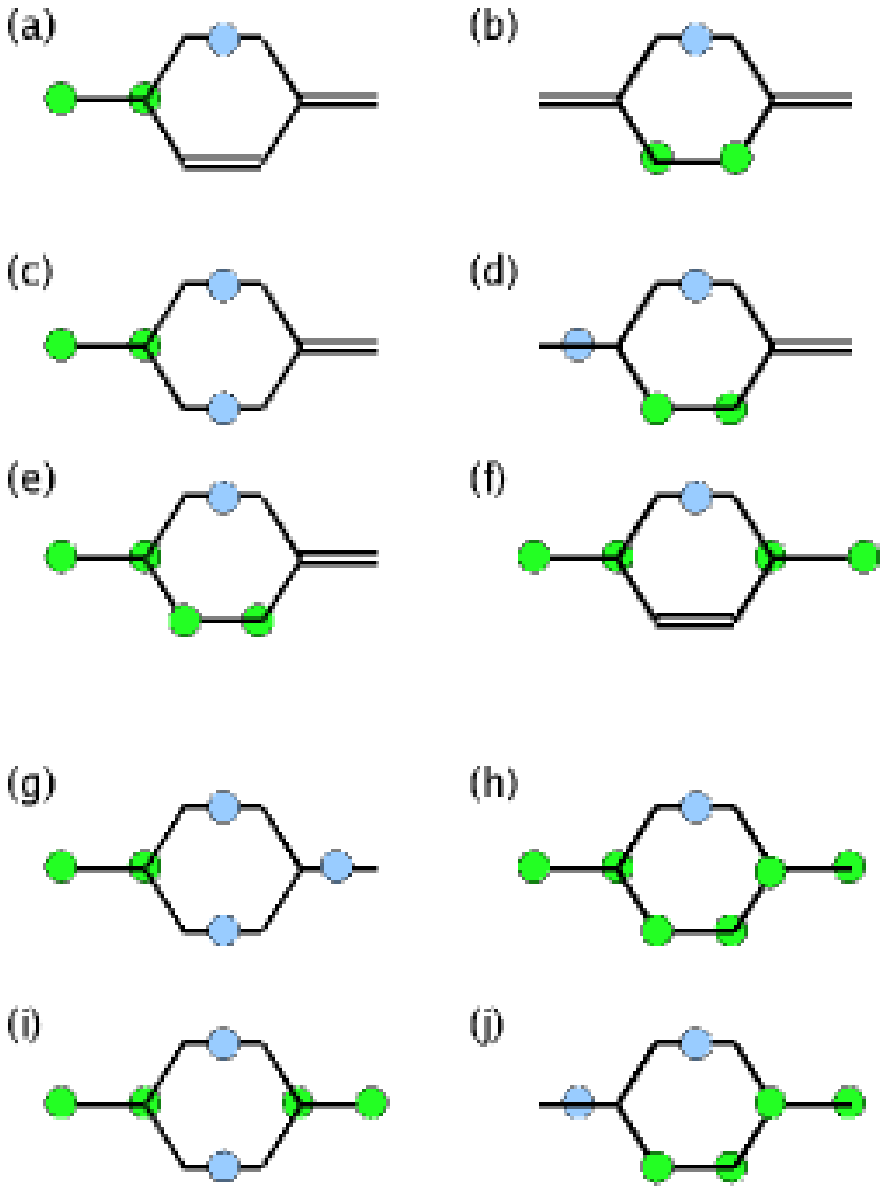}
\caption{A sketch of functionalization of graphene by oxygens
(blue circles) and hydroxyl groups (green circles).}
            \label{fig4}
 \end{center}
\end{figure}

Let us consider now a general case of functionalization of
graphene by both hydroxyl groups and oxygen atoms. Typical
combinations are shown in Fig. \ref{fig4}. Total energy
calculations demonstrate that for all the combinations under
consideration the chemisorption energy per hydroxyl group is 60
meV lower, and per oxygen atom 30 meV lower than for the pure
cases with the same degrees of coverage. Thus, mixed coverage is
energetically favorable in these cases (coverage between 25\% and
75\%) diminishing the energy of both O and OH groups. For coverage
less than 25\% the chemisorption energy for hydroxyl groups turns
out to be lower than for oxygen, therefore one can conclude that
GO with 25\% coverage contains only OH groups whereas some oxygen
atoms can appear only as edge groups. Optimal configurations for
25\%, 50\%, and 100\% coverage are shown in Figs. \ref{fig4}a, e
and c, and h, respectively. One should stress that a structure
with staggered stripes of $sp^2$ and $sp^3$ carbon atoms is formed
at 50\%, in agreement with the previous works$^{21,22}$. For the maximal
coverage, as well as for the case of OH groups only, carbon-carbon
distances exceed 1.54 \AA ~which makes, again, 100\% coverage less
favorable than 75\% one. The most stable configuration of GO is
presented in Fig. \ref{fig1}c.

As a result, one can suggest the following chemical formulas for
GO with various coverage: 25\% - C$_8$(OH)$_2$, 50\% -
C$_8$(OH)$_2$O, and 75\% - C$_8$(OH)$_4$O. They are rather close
to the formulas suggested by experimentalists and discussed in the
Introduction. Minor discrepancies can be related with the presence
of some small amount of carboxyl and carbonyl groups, as well as
atomic oxygen adsorbed at the edges of GO, as was discussed in
detail within the model proposed by Lerf et al$^{19,20}$.

\begin{figure}[ht]
 \begin{center}
   \centering
\includegraphics[width=5.2 in]{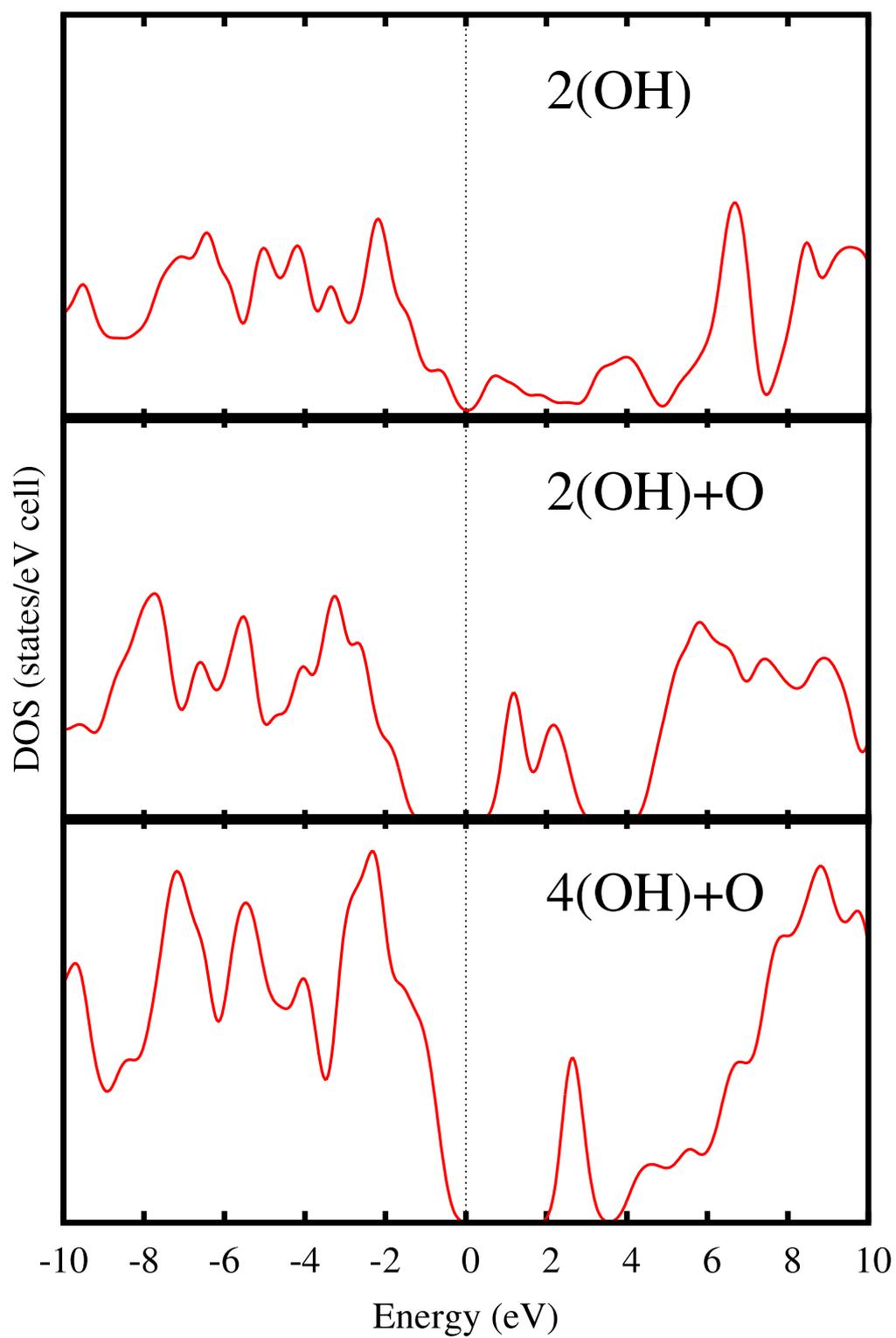}
\caption{Electronic densities of states for the most stable
configurations at various degrees of coverage. Number of hydroxyl
groups and oxygen atoms per C$_8$ is shown.}
            \label{fig5}
 \end{center}
\end{figure}

Electron densities of states for GO are presented in Fig.
\ref{fig5}. One can see that the energy gap varies between 2.8 eV
and 1.8 eV at the decrease of coverage from 75\% to 50\%. At
further reduction of GO it becomes conducting, according to our
calculations. It seems to be in agreement with the available
experimental data$^{4,25,37}$.

\begin{figure}[ht]
 \begin{center}
   \centering
\includegraphics[width=5.2 in]{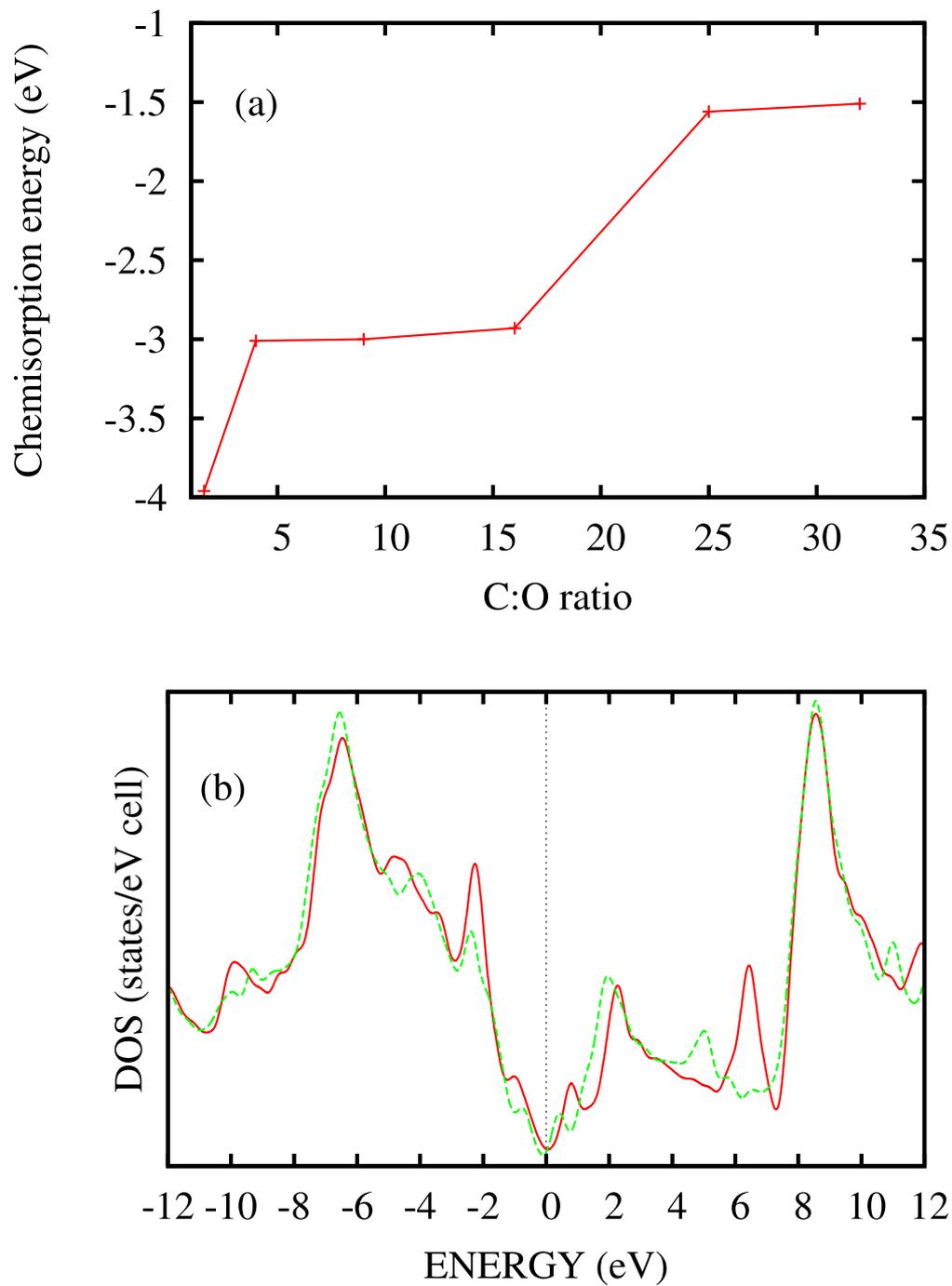}
\caption{ (a) Chemisorption energy of OH groups as
a function of C:O ratios;
(b) Total densities of states per atom for C:O ratios
16:1 (solid red line) and 32:1 (dashed green line).}
            \label{fig6}
 \end{center}
\end{figure}

The chemisorption energy difference per group for 25\% and 75\%
coverage is less than 1 eV (Fig. \ref{fig3}a) which explains a
possibility of partial reduction of GO, both thermally and
chemically.
Actually, the carbon to oxygen ratio 4:1 considered above is a bit
larger than experimental values for strongly reduced GO$^{14,25,38,39}$
and almost twice larger than the maximal ratio 10:1$^{37}$. To study dependence of the
chemisorption energy on the C:O ratio we have performed
calculations for the cases of two hydroxyl groups (see Fig. \ref{fig2}f)
per supercells containing 8, 18, 32, 50, and 72 carbon atoms, the
latter case corresponding to the C:O ratio 32:1. The
computational results are presented in Fig. \ref{fig6}a. One can see that
the chemisorption energy is weakly concentration-dependent between
the ratio values 4:1 and 16:1 whereas for smaller concentrations of
hydroxyl groups it decreases roughly twice, between 16:1 and 25:1.
A weakening of chemical bonding between OH groups and graphene
at small concentrations (C:O ratio from 25:1 in comparison with 16:1)
is connected with essential changes of the electronic structure.
At very small concentrations, the latter is more similar to that of
pure graphene (see Fig. \ref{fig6}b).
It can be caused by the changes of distances between OH groups
which is 17 \AA ~for the carbon-to-oxygen ratio 16:1. It was shown
at the simulation of hydrogenization of graphene$^{26}$ that
typical radius of interaction between the hydrogen atoms is about
8 \AA, and the defects can be considered as independent ones for
larger distances. For the case of OH groups, the distortions of
graphene sheet is larger than for the case of hydrogen and
therefore interaction between them is still essential for the
ration 16:1 whereas for smaller concentrations the hydroxyl groups
can be considered as almost non-interacting. In real experimental
situation where the ratio 10:1 has been reached$^{37}$
finite-size effects of the GO flakes can be important. Indeed, the
size of these flakes is smaller than for the case of graphene$^{9,23}$
and various groups can be chemisorbed at the
edges. Also, the flakes can contain various topological defects$^{10,23}$
which can also change local chemisorption energy.

\begin{figure}[ht]
 \begin{center}
   \centering
\includegraphics[width=5.2 in]{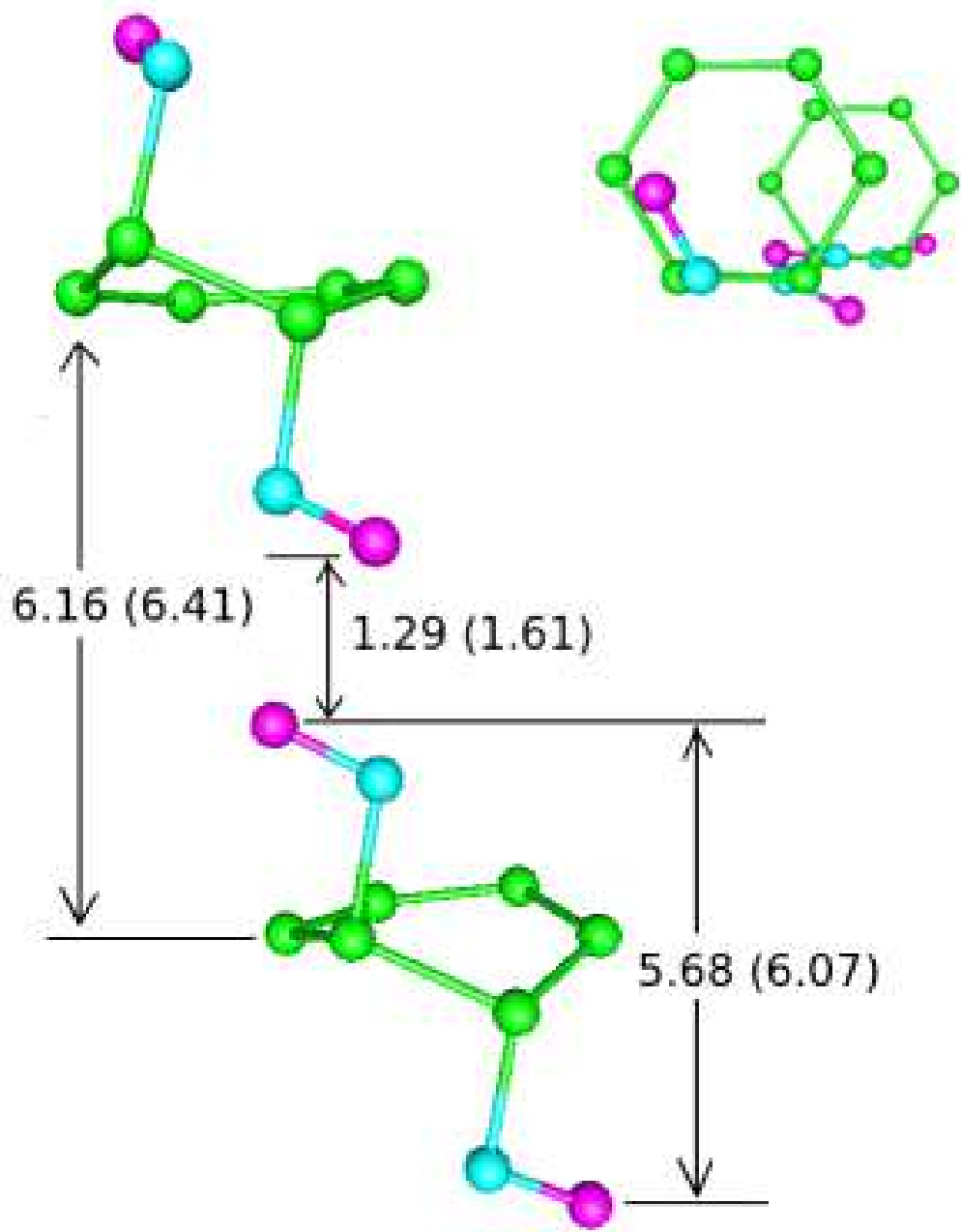}
\caption{Optimized geometric structures of strongly reduced GO.
Numbers are distances, in \AA, for the periodic structure (and for
bilayer in parentheses). Right upper corner: a top view. Carbon,
oxygen and hydrogen atoms are shown in green, blue, and violet,
respectively.}
            \label{fig7}
 \end{center}
\end{figure}

At last, let us discuss the cases of bilayer and periodic
(graphite-like) GO. To this aim we have carried out calculations
for corresponding structures with 25\% coverage by hydroxyl
groups. We consider the Bernal
(AB) stacking, similar to pure graphite, which was observed also
in GO$^{16}$. The optimized structure is shown in Fig.
\ref{fig7}. The width of the layer was found to be, in both cases,
about 7 \AA, as well as interlayer distances, which seems to be in
a good agreement with experimental data$^{38,40,41}$.
To calculate interlayer coupling energy per carbon atom
we have computed the energy differences between single layer and
periodic structure. For the case of pure graphite, it equals 32
meV, in a good agreement with the experimental value 35 meV$^{42}$.
For the periodic GO structure and for GO bilayer, the
corresponding values turn out to be 17 meV and 6 meV,
respectively. This decrease of the energy explains possibility of
easy exfoliation in GO$^{9,10}$. Due to weakness of
interlayer coupling, the electronic structure of GO is almost
identical for single layer, bilayer, and periodic structure, in
contrast with the cases of pure graphene$^{43}$.

\section{Conclusions}

To summarize, we have proposed a model structure of GO which seems
to be consistent with all known experimental data. We have
demonstrated, in particular, that (i) 100\% coverage of GO is less
energetically favorable than 75\% (ii) functionalization by both
oxygen and hydroxyl groups is more favorable for coverage than
25\% than by hydroxyl groups only (iii) a reduction of GO from
75\% to 6.25\% (C:O ratio 16:1) coverage is relatively
easy but further reduction seems to be rather difficult,
and (iv) GO becomes conducting at the
coverage 25\%, being an insulator for larger coverage.

{\bf Acknowledgment} ~~
The work is financially supported by Stichting voor
Fundamenteel Onderzoek der Materie (FOM), the Netherlands.

{\bf Supporting Information Available:} Cartesian coordinates for
all species and values of calculated formation energies. This material
is available free of charge via the Internet at http://pubs.acs.org.

\end{document}